\documentclass[aps,prl,twocolumn,superscriptaddress,showpacs,
floatfix,nofootinbib]{revtex4-1}

\usepackage{dcolumn}
\usepackage{bm}
\usepackage{color}
\usepackage{amssymb}
\usepackage{amsmath}
\usepackage{graphicx}
\usepackage{float}
\usepackage{natbib,hyperref}

\begin{document}
\title{Predicting Energies of Small Clusters from the Inhomogeneous Unitary Fermi Gas }

\author{J. Carlson}
\author{S. Gandolfi}
\affiliation{Theoretical Division, Los Alamos National Laboratory, Los Alamos, NM, 87545}

\date{\today}
\begin{abstract}
We investigate the inhomogeneous unitary Fermi gas and use the long-wavelength
properties to predict the energies of small clusters 
of unitary fermions trapped in harmonic potentials. 
The large pairing gap and scale invariance place severe
restrictions on the form of the density functional. We determine
the relevant universal constants needed to constrain the functional from calculations of the bulk 
in oscillating external potentials.  Comparing with exact Quantum Monte Carlo 
calculations, we find that the same
functional correctly predicts the lack of shell closures
for small clusters of fermions trapped in harmonic wells as well
as their absolute energies.  A rapid convergence to the bulk
limit in three dimensions, where the surface to volume ratio is quite large,
is demonstrated.  The resulting 
functional can be tested experimentally, and is a key ingredient
in predicting possible polarized superfluid phases and the properties of the
unitary Fermi gas in optical lattices.

\end{abstract}

\pacs{71.15.Mb,03.75.Ss,67.85.Lm}

\maketitle

The properties of the homogeneous unitary Fermi gas have been the subject of
intense experimental and theoretical  study over the 
last decade~\cite{Giorgini:2008,Randeria:2013},
including studies of the equation of state~\cite{Ku:2012,Navon:2010}, 
the spin and density response at high momentum transfer~\cite{Hoinka:2012},
and other properties related to the contact parameter~\cite{Hoinka:2013}.  
More recently the inhomogeneous
gas is receiving considerable attention, in particular
studies of the three- to two-dimensional transition\cite{Sommer:2010},
fermions in optical lattices\cite{Bloch:2008,Ma:2012, Pilati:2014}, 
and studies of small clusters of cold fermions at 
unitarity\cite{Zurn:2013,Wenz:2013}. In this paper we study the 
inhomogeneous unitary Fermi gas in three
dimensions and show that its properties uniquely determine the
properties of even very small clusters of trapped fermions.

We show that a relatively simple density functional,
obtained by fitting the properties of the unitary Fermi gas in periodic potentials, can 
accurately reproduce the energy of small
clusters of atoms confined in an harmonic trap.  This outcome is
not only relevant for cold atoms, but is of high interest for nuclear
physics, inhomogeneous  superconductors, properties of atoms in optical
lattices, and in  material science (see for example
Ref.~\cite{Ma:2012}).

In this work we study the system described by the Hamiltonian:
\begin{equation}
H = \sum_i e_i(q) +  
g\,\sum_{k,p,q} a_{\uparrow k+q}^\dagger a_{\downarrow p-q}^\dagger 
             a_{\uparrow k} a_{\downarrow p} + V_{ext},
\label{eq:unitaryh}
\end{equation}
where the dispersion is either $e_i(q) = q^2/(2m)$, or other improved lattice actions~\cite{Carlson:2011},
 and
the potential strength $g$ is tuned to describe infinite scattering length
in the two-particle system. The resulting system is the scale invariant
Unitary Fermi gas (UFG), characterized by universal constants including
the ratio $\xi=E/E_{FG} \approx 0.37$~\cite{Carlson:2011,Ku:2012} of 
the UFG to free Fermi gas energy, that is density independent.

   The UFG has many special properties, including a very large
ratio of pairing gap to Fermi energy,
$\Delta/ E_F \approx 0.45(5)$ \cite{Carlson:2005,Carlson:2008,
Schirotzek:2008},
and two universal constants that describe all the long-wavelength
excitations~\cite{Son:2006}.
Pairing gaps of the scale of the Fermi energy imply that in the ground state
single fermions cannot propagate over long distances, and thus moderate
sized clusters of fermions will not exhibit the large
shell effects typically found in atoms or nuclei.
Large pairing gaps in neutron-rich atomic nuclei have been used to
describe shell quenching, or the lack of closed shell gaps in 
energy versus particle number characteristic of atoms or of nuclei
with $N \approx Z$~\cite{Dobaczewski:1994}. In addition, the strong pairing makes these systems
behave similarly to Bose gases, 
where shell effects that are usually evident for Fermions in finite boxes are 
totally absent for the UFG~\cite{Forbes:2012}.
The unitary limit provides a clean 
simple experimental system that can further our understanding of the
transition from small systems to the bulk.
The energies of a small number of fermions in one dimension have
recently been studied experimentally to determine this transition to
bulk behavior~\cite{Wenz:2013}.
In three dimensions one would expect the transition to 
require much larger systems because of the larger surface to
volume ratio.

The two universal constants describing low-energy phonon
excitations and the susceptibility to long-wavelength
oscillations can be calculated in the bulk.
The scale invariance
and the large pairing gap place severe restrictions on the density functional
of the UFG, i.e. in a gradient expansion all terms must have
the same dimension as the free Fermi gas energy density, or one over
length to the fifth power, or $\rho^{5/3}$~\cite{Son:2006}.  The susceptibility as a function of
wavelength then completely determines the density functional, and
can be used to make unique predictions for small clusters.

In order to calculate the above mentioned properties,
we perform both Diffusion Monte Carlo (DMC) and Auxiliary Field
Monte Carlo (AFMC) calculations of the inhomogeneous UFG.
Both extract the ground state from a Monte Carlo evaluation of
$| \Psi_0 \rangle = \exp [ - H \tau ] | \Psi_T \rangle$.
The AFMC calculations are exact for a specific lattice size since they
do not suffer from a sign problem~\cite{Chen:2004,Lee:2005,Bulgac:2006}.  
The method involves a branching Markov chain
Monte Carlo where the states are evolved through a fluctuating auxiliary field.
The algorithm is described in \cite{Carlson:2011}, the only addition is that we
modify the BCS trial state $| \Psi_T \rangle$ multiplying the
pairing terms $\phi (r_{ij})$ by a
product of two single-particle functions $\Phi (r_1 ) \Phi (r_2 )$. This
additional importance sampling lowers the statistical errors considerably 
for strong external fields. The Markov chain Monte Carlo is implemented
as a branching  walk, allowing
us to extend the calculations to extremely low temperatures
with modest variance.  We have made calculations on cubic lattices
of size $L^3$, with $L$=16, 20, 24, and have also used two different
dispersion relations $e_i (q)$~(Eq. \ref{eq:unitaryh})
 to eliminate effective-range and other effects from
finite lattice spacing. 

\begin{figure}[t!]
\vspace{-0.6cm}
\includegraphics[width=0.45\textwidth]{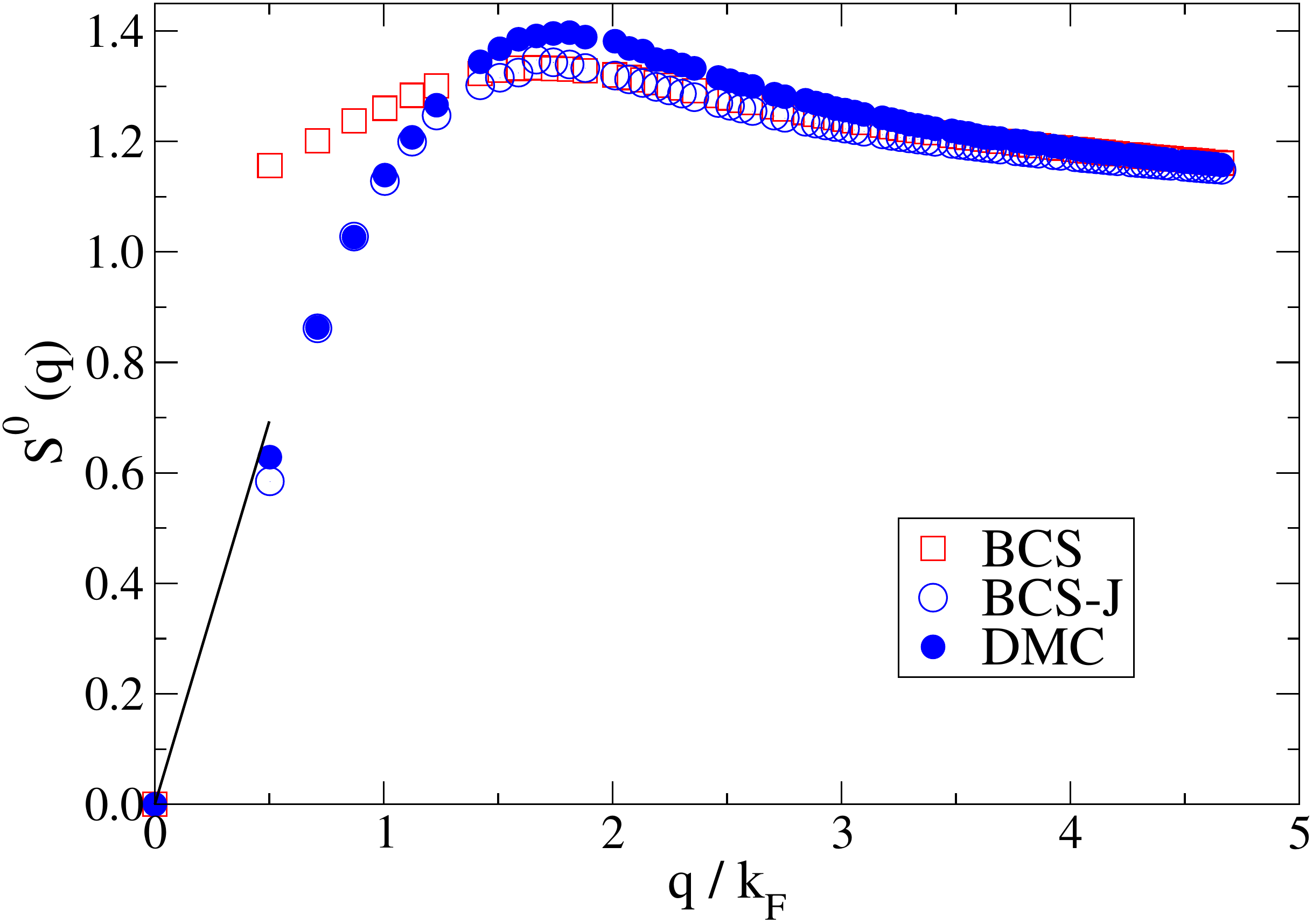}
\vspace{-0.2cm}
\caption{(Color online) Static structure function for the unitary Fermi gas.
The open squares are the BCS result, and the open circles 
are for the BCS-J variational trial function. The filled circles
are the results of the DMC calculation (see text).}
\vspace{-0.6cm}
\label{fig:sq}
\end{figure}

The DMC calculations provide accurate
upper bounds to the ground state energy, but can be performed in
the extremely dilute limit compared to the lattice calculations.
Other operators can also be easily computed~\cite{Gandolfi:2011}.
The DMC calculations use optimized BCS variational wave functions
described in~\cite{Gandolfi:2011,Sorella:2001}.  
We add long range correlations to the BCS wave function to improve
the description of the low energy physics.
We first consider the static structure factor
$ S^0 (q) = \langle 0 | \rho^\dagger ({\bf q})
\rho ({\bf q}) | 0 \rangle$ 
with 
$\rho ({\bf q}) = \sum_i \exp [ i {\bf q} \cdot {\bf r}_i ].$
The energy and inverse energy weighted sum rules are denoted 
$S^1 (q)$ and $S^{-1} (q)$, respectively.
Figure \ref{fig:sq} shows $S^0 (q)$ obtained from the optimized
BCS wave function without (BCS) and with (BCS-J) long-range correlations and the
results of our DMC calculations.  The 
trial wave function
including long-range Jastrow correlations~\cite{Reatto:1967,Vitali:2011} is:
\begin{equation}
\Psi_{BCS-J}=\prod_{i<j}\exp\left[\gamma\sum_n\frac{\exp(-\lambda|{\bf q_n}|)}{|{\bf q_n}|}
\exp(-i{\bf q}\cdot{\bf r}_{ij})\right]\Psi_{BCS} \,,
\end{equation}
where $\Psi_{BCS}$ is the correlated BCS wave function with short-range
Jastrow correlations used in previous 
calculations~\cite{Gandolfi:2011}, and $\gamma$ and $\lambda$ are 
variational parameters.
The final DMC energies
are independent of the choice of these correlations.
In Fig.~\ref{fig:sq} the solid line at low $q$ is the phonon dispersion  $S(q)=(q/k_F)\sqrt{3/\xi}/2$, using the convention $\hbar^2/(2m) = 1$.
A small deviation from the linear dispersion is apparent at $q/k_F=0.5$,
where we find $S^1(q)/(k_F^2 S^0(q))=0.402(5)$.

The static density susceptibility at low $q$ , given by the inverse
energy-weighted sum rule $S^{-1} (q)$,
determines the long-wavelength behavior of
the density functional.  
The first gradient-squared correction to the local density approximation
is unique, we rewrite the density functional
obtained in \cite{Rupak:2009} as:
\begin{equation}
{\cal E}_2 \ = \ \int V_{ext} (r) \rho (r) \ + \xi \frac{3}{5}
( 3 \pi^2 )^{2/3} \rho^{5/3} \ + c_2 \ \nabla \rho^{1/2} \cdot \nabla \rho^{1/2}.
\label{eq:densitygradient}
\end{equation}
The first two terms are the local density approximation (LDA), ${\cal E}_2$
depends on only one additional parameter $c_2$,
and remaining terms are of order $q^4$ or higher.  This
form makes explicit the connection between the BCS and BEC limits,
in this notation $c_2 = 0.111$ in the extended Thomas Fermi model
for free fermions, and $c_2 = 0.5$ in the extreme BEC limit. The
gradient term is exactly the same as for the Gross-Pitaevskii (GP)
equation used to describe Bose-Einstein condensation \cite{Dalfovo:1999},
in this case condensation of pairs.
This parameter is important in GP treatments
of the dynamics in the unitary gas, including soliton and vortex
dynamics~\cite{Yefsah:2013,Bulgac:2014},
and to possible inhomogeneous superfluid (LOFF) phases
of the polarized system~\cite{Bulgac:2008}.

To determine the static response at unitarity, we calculate the energy
of the system in an external potential $V_{ext} = V_0\  E_F\  
\sum_i\cos ( {\bf q} \cdot {\bf r}_i)$, with $E_F \equiv (3 \pi^2 \rho)^{2/3}$. 
Initially we choose $V_0 = 0.25$ and $q/k_F = 0.5$
to remain in the low momentum limit and yet have a large enough
energy difference with the uniform system to have a statistically
meaningful result.
From DMC and AFMC results we obtain
$c_2 = 0.30(5)$, extracted from a standard
bosonic  DFT calculation that minimize ${\cal E}_2$.
The extracted $c_2$ is larger than
that obtained in the $\epsilon$ expansion\cite{Rupak:2009}, and
between the known values in the BCS and BEC regimes. 
The value $c_2 < 0.5$ implies slower (larger effective mass) dynamics 
than a simple GP treatment with $c_2 = 0.5$ from two paired fermions.
The extracted $c_2$ and the $S^0 (q)$ can be used to fix the two low-energy
parameters in the effective field theory\cite{Son:2006} and to
constrain additional terms in more sophisticated density functionals
such at the superfluid local density approximation (SLDA)\cite{Bulgac:2007,Forbes:2012} that include fermionic degrees of freedom
and can treat polarized systems.

\begin{figure}[t!]
\begin{centering}
\includegraphics[width=0.45\textwidth]{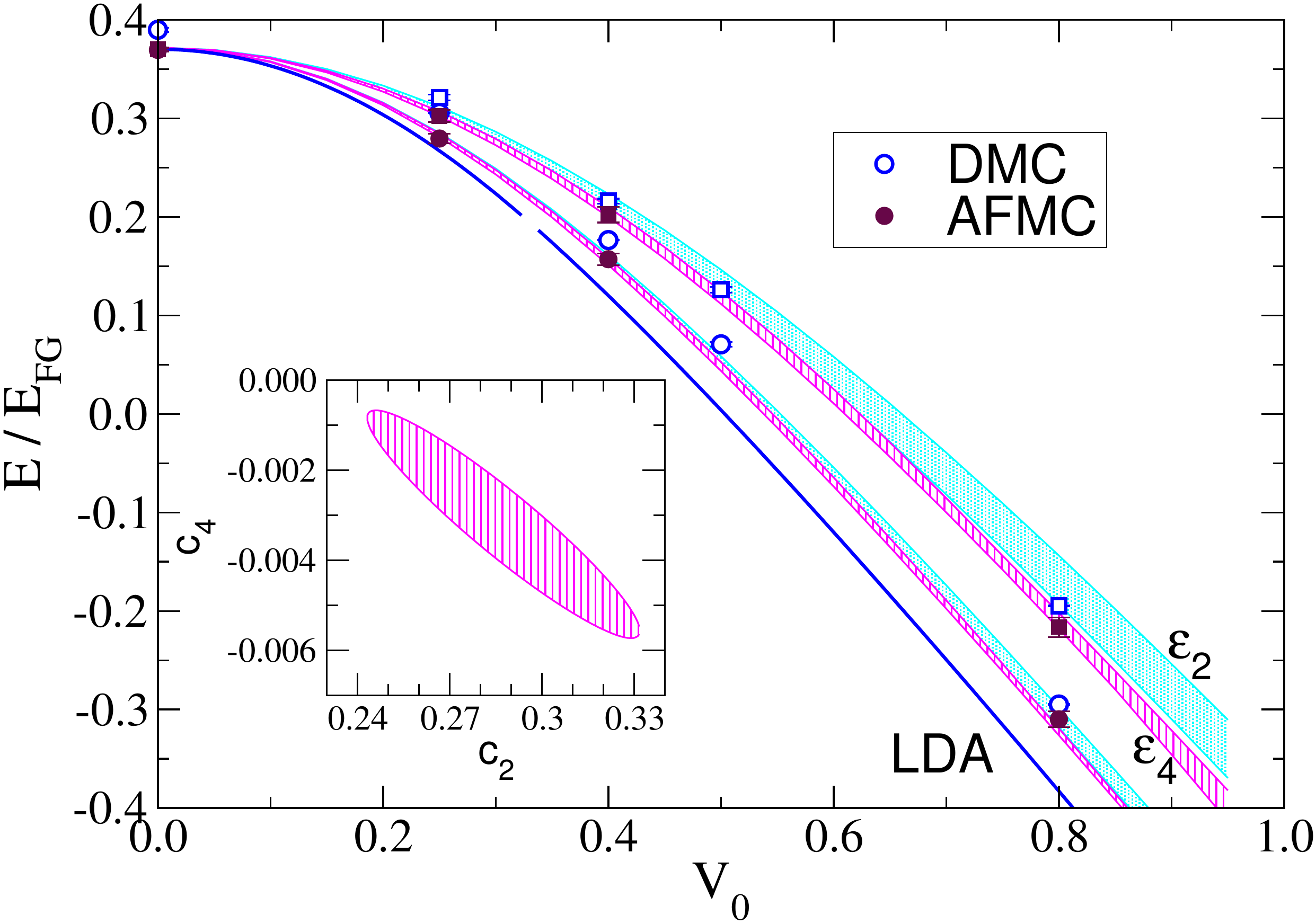}
\end{centering}
\vspace{-0.3cm}
\caption{(Color online) Energy of the unitary Fermi gas in a 
periodic potential versus strength
of the interaction for $q = k_F/2$ (lower curves) and $q=k_F$ (upper curves). Quantum Monte Carlo calculations are shown as symbols. The bands are density
functional results for ${\cal E}_2$ using  $c_2 = 0.30(5)$  and for
${\cal E}_4$ with $c_2$ and $c_4$ extracted from fits to all the bulk QMC
data. See the text for details. The error ellipse obtained for $c_2$ and $c_4$ from the fit 
is shown in the inset.}
\vspace{-0.5cm}
\label{fig:staticfit}
\end{figure}

For larger gradients (larger $V_0$) higher-order terms become important.
We have calculated the energy of 66 particles in periodic boundary
conditions for a range of $V_0$ for $q/k_F = 0.5$ and $q/k_F = 1$, shown in Fig.~\ref{fig:staticfit} as
lower and upper symbols and bands, respectively.
The blue solid line
indicates results expected in the local density approximation without
gradient terms, entirely determined by $\xi$.  
The break in this line represents the point at which
the density separates into quasi two-dimensional sheeets.
The results of the DMC and AFMC calculations are shown
as open and closed symbols respectively.  

Using the coefficient $c_2$ obtained for weak external fields,
the QMC calculated energies for $q = 0.5\ k_F$ are well reproduced by this
density functional for the whole range of $V_0$ (lower solid band). 
This simple density
functional is expected to work very well for systems where 
$|\nabla \rho / (k_F \rho)| << 1$ everywhere.  
In Fig.~\ref{fig:staticfit} it is evident that for the 
larger $q = k_F$, the ${\cal E}_2$ density functional begins to
fail, particularly at larger $V_0$. In this region the higher order
gradient corrections are becoming important.

The first correction to the simple gradient density functional ${\cal E}_2$
(Eq. \ref{eq:densitygradient}) is of order $q^4$ \cite{Rupak:2009}.
It is natural to find the energies at higher momenta smaller than those
given by ${\cal E}_2$, this behavior would be expected based on the typical
roton-phonon spectrum~\cite{Ku:2012,Taylor:2009}.  Using
the scale invariance of the density functional and  a Negele-Vautherin\cite{Negele:1972}
expansion for the density functional in terms of gradients, we add
another term
\begin{equation}
{\cal E}_4 \ =  \ {\cal E}_2 \ + \ c_4 \frac{\nabla^2 \rho^{1/2} \nabla^2 \rho^{1/2}}{\rho^{2/3}} \,,
\end{equation}
with the same dimensions as ${\cal E}_2$.
This additional term is attractive $(c_4 < 0)$ since the quasiparticle spectrum
lies lower than the simple linear behavior with increasing $q$.  

We perform a simultaneous fit of $c_2$ and $c_4$ in ${\cal E}_4$ to all the
AFMC data to obtain the error ellipse shown in the inset of Fig.
\ref{fig:staticfit}. For each pair of values $c_2$ and $c_4$
a standard DFT calculation of the density is first performed setting $c_4 = 0$,
then the energy contribution from the $c_4$ term in ${\cal E}_4$ is calculated
perturbatively from this density distribution.
Since the $q^4$ term in ${\cal E}_4$ term is attractive, we must evaluate it
perturbatively as it is unstable to high-frequency oscillations. Higher-order
terms including those associated with the contact would stabilize the system
\cite{Son:2010}.

The extracted error ellipse for these parameters shows
a strong correlation since a larger value of
$c_2$ requires a more attractive value of $c_4$.
The solid and vertical hatched regions give the error bands
for ${\cal E}_2$ and ${\cal E}_4$, respectively.
The ${\cal E}_4$ density functional provides an excellent fit to all the
data, with a $\chi^2$ per degree of freedom near one.
The width of the bands in the main figure 
represent varying the coefficients within
the quoted uncertainties (the inset ellipse for ${\cal E}_4$).

\begin{figure}
\includegraphics[width=0.40\textwidth]{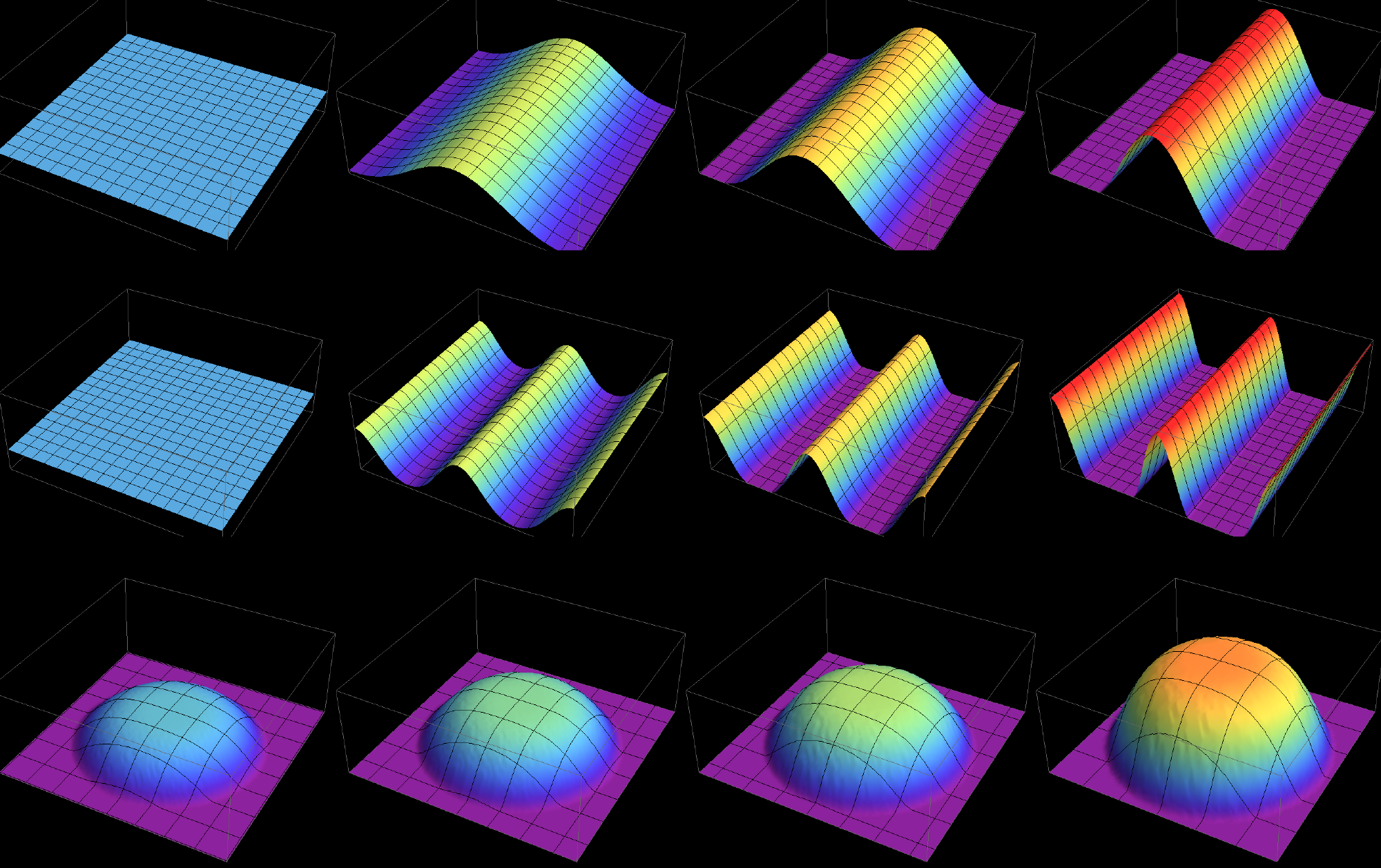}
\caption{(Color online) Densities of the unitary Fermi gas in external potentials of frequency $k_F/2$ (upper row) and 
$k_F$ (middle row) for potential strengths $V_0$ = 0, 0.25, 0.4, and 0.8 from left to right.
The lower row shows the predicted density distributions (in the z=0 plane) 
for systems of 8, 14, 30, and 50 fermions (left to right) in a harmonic trap.
Scale invariance requires the energies depend only upon the shape of the
density distribution, except for an overall scale of $\rho^{2/3}$.}
\vspace{-0.6cm}
\label{fig:dens}
\end{figure}

The density functional can then be used to predict the densities
of inhomogeneous matter and properties of small numbers
of fermions trapped in harmonic wells.  Observing the densities in
an external field should be 
an accurate way to measure the coefficients in the density functional.
The densities for both inhomogeneous
matter and small trapped systems are shown in Fig. \ref{fig:dens}.
The upper two rows illustrate the 
transition from three towards two dimensional systems with 
increasing $V_0$ for external 
potentials of momenta $k_F/2$ and $k_F$, and the bottom row
shows the densities of small systems trapped in a 
harmonic potential.

\begin{figure}
\vspace{-1.0cm}
\begin{centering}
\includegraphics[width=0.45\textwidth]{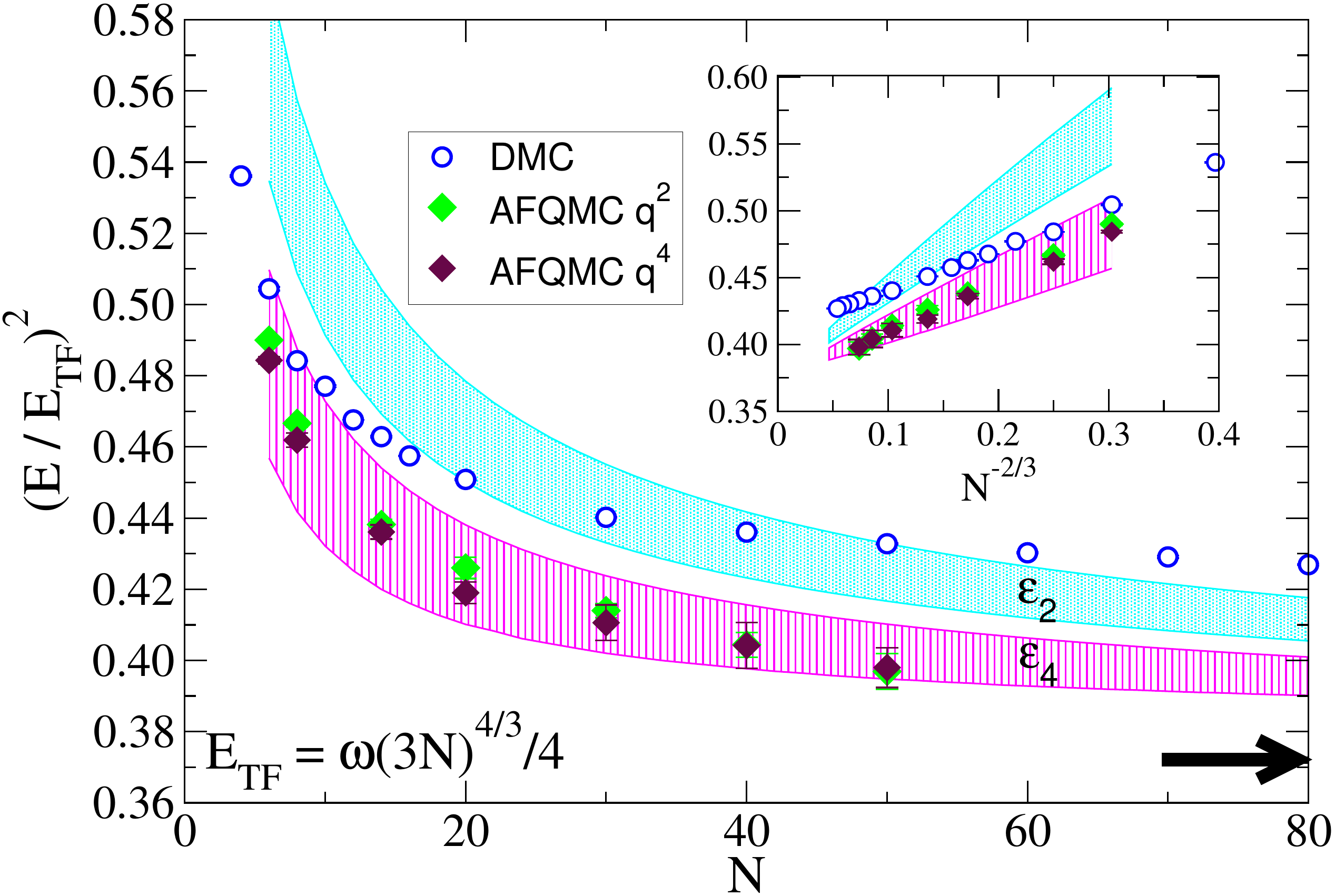}
\end{centering}
\vspace{-0.7cm}
\caption{(Color online) Ground-state energy 
($E(N)/E_{TF})^2$ of trapped fermions at unitarity vs. particle number N.
Present DMC calculations are shown as open circles, and AFMC calculations
as diamonds. Density functional results ${\cal E}_2$ and ${\cal E}_4$
are shown (see text). The inset shows the extrapolation of the same data
to the bulk (large N) limit.}
\vspace{-0.6cm}
\label{fig:hoplot}
\end{figure}

To check the predictions for trapped fermions, 
we calculate systems of fermions at unitarity in a harmonic
trap from 4 to 80 particles (Fig. \ref{fig:hoplot}). 
The square of the ratio of the energy at unitarity
to the Thomas Fermi energy for free fermions,
$E_{TF} = \omega (3N)^{4/3}/4$, is plotted as a function of the number
of particles.  This ratio should approach the bulk (LDA) limit as the size
of the system increases.
The DMC results are shown
as blue open circles in the figure, and the AFMC results are shown as diamonds.
For $N > 8$, both our DMC and AFMC results are significantly lower than those obtained previously
by Endres, et al.\cite{Endres:2011},
Blume, et al.\cite{Blume:2007}, Chang and Bertsch\cite{Chang:2007}, and by 
Mukherjee and Alhassid\cite{Mukherjee:2013}.
The AFMC calculations
extend to much lower temperature T than previous lattice calculations,
and are averaged from $ \omega/T  \approx 4-10$.

The DMC calculations include a more sophisticated trial wave function
than used previously. It includes pairing both in a single-particle
orbitals as typically used in atomic nuclei
and pairing based upon the local density approximation. 
The variational wave function for the 
system in the trap has pairing orbitals
with the following form:
\begin{eqnarray}
\Phi({\bf r}_1,{\bf r}_2)&=&
\left[\sum_i d_i \phi_{n_i}^{HO}(\alpha_i{\bf r}_1) \phi_{n_i}^{HO}(\alpha_i{\bf r}_2)\right]
e^{-(\gamma_1+\gamma_2) R^2}  
\nonumber \\
&+&\beta[k_F(R)\,r]e^{-\frac{m\omega}{2\hbar}R^2}(1-e^{-\gamma_2 R^2}) \,.
\end{eqnarray}
where $n_i$ are HO quantum numbers of the $i$-th state, $R=|{\bf r}_1+{\bf r}_2|/2$, 
$r=|{\bf r}_1-{\bf r}_2|$, and the function $k_F(R)$ is the local momentum 
as a function of the center of mass of the pair: $
k_F(R)=\left[\frac{1}{\hbar\xi}(\xi E_{F}-\omega^2 R^2/2)\right]^{1/2}$, and 
the function $\beta(r)$ has the same form of Ref.~\cite{Gandolfi:2011}.
The variational parameters $d_i$, $\alpha_i$ and $\gamma_i$ are optimized,
and simulations at different effective ranges to extract the zero-range
limit. If we simplify our calculations to the simple trial function
used in \cite{Blume:2007} and \cite{Chang:2007}, we reproduce their
higher energies.

The AFMC lattice calculations are exact but subject to finite lattice size errors.  Two sets of AFMC calculations are shown, one using the $q^2$ dispersion
relation, the other using the $q^2 + q^4$ dispersion 
discussed in \cite{Carlson:2011}.
The finite-size energy correction for the $q^2$ dispersion is 
proportional to the effective range and to the lattice
spacing, it is given by:
\begin{equation}
\delta E_{HO}(N) / E_{TF} \ = \ -\frac{2048 N^{1/6} \omega^{1/2} S r_e }{525 \times  3^{5/6} \pi \xi^{3/4}}.
\end{equation}
Numerically this yields $\delta E(N)/E_{TF} \approx 0.0388\ \omega^{1/2} N^{1/6}$ for $S = 0.12$, 
$\xi = 0.37$, and $r_e = 0.337 a$ ($a$ is the lattice spacing), the correction is approximately a 2\% lowering of the QMC energy.
The value of $S$ is extracted from Refs.~\cite{Carlson:2011,Forbes:2012b}.
Similar corrections have been applied in the bulk, they are
significantly smaller than the statistical errors.

Similarly the $q^4$ propagator requires a correction from pairs of finite momentum, which for small lattice sizes lowers the energy from the continuum behavior. Calculating the energy of a pair with finite momentum yields a
correction $\delta E(N)/ E = \zeta^2 a^2 (5/6) \langle {\bf K}^2 {\bf k}^2 \rangle$,
where $\zeta = 0.16137$ is the coefficient of the $k^4$ term in the
propagator, 
and $ {\bf k}^2 $ and $ {\bf K}^2$ are the average square
momenta of particles and pairs respectively. The former can be estimated
from the simulation and the latter from the calculated total energy using
the virial theorem. In this case the correction yields an approximately
1\% increase in the energy. This correction is numerically consistent
with zero for homogeneous systems as shown by 
calculations of two species
of unequal mass~\cite{Carlson:2012,Gezerlis:2009}. The
two sets of AFMC energies calculated with different dispersions
agree within error bars.  
The corrections for the periodic external potential are much smaller than 
the error bars since the external interaction confines the system in
only one dimension.

The QMC results for small clusters are compared with the predictions from the
two different density functionals ${\cal E}_2$ 
and ${\cal E}_4$ in Fig.~\ref{fig:hoplot}.
In the local density approximation the ratio of squared energies is 
a constant $\xi$ for any N, the arrow indicates the bulk value of $\xi$
applicable in the large N limit.  The results for ${\cal E}_2$
are shown as the upper solid band, and the 
predictions from ${\cal E}_4$ are shown as the lower vertical hatched band. 
This density functional provides an excellent description of the small
trapped systems, the $c_4$ term is much more important in this case.

As we can see in Fig.~\ref{fig:hoplot} our calculations yield no significant shell
effects or breaks in the curve of $E/ E_{TF}$ curve
versus the number of particles.
In the BCS limit there would be
sharp breaks of the energy with particle number, with closed shells at
$N = 2, 8, 20, 40,...$ for a harmonic oscillator external potential.
Shell closures are a natural expectation for many fermionic systems,
even those  with significant pairing like atomic nuclei or inhomogeneous
neutron matter~\cite{Gandolfi:2011b}.   
In the unitary Fermi gas, however, the shell breaks appear quite small,
further justifying the density functional in terms of the local density
and its gradients.  This is to be expected for large systems, where
the coherence length is much smaller than the system size. Even for small
systems, it would appear that unpaired fermions cannot propagate
significantly.
This physics has a natural analogue in neutron-rich
atomic nuclei, where the pairing gaps play an increasingly important
role compared to shell gaps as the number of neutrons increase.

In summary, we find find that the density functional of the Unitary Fermi gas
is strongly constrained by the scale invariance of the system and the
large pairing gap for single particle excitations.
A simple density functional 
with two parameters beyond the LDA
can describe the energies of inhomogeneous systems over a very wide range
of external potentials.
We have extracted the universal values of the density functional
coefficients from calculations in a one-dimensional periodic 
potential, and we find that
even very small trapped systems can be successfully predicted by
this density functional. These small trapped systems
show no evidence of significant shell closures as would be expected
for most Fermionic systems.
Among other applications, this density functional could be tested by
predicting properties of the UFG in optical lattices, and compared with
experiments.

\acknowledgments{
We would like to thank Kevin E. Schmidt, Shiwei Zhang and Sebastiano Pilati for stimulating
discussions. 
Computer time was provided by an INCITE allocation and by Los
Alamos Institutional Computing. 
This research used also resources of the National Energy Research
Scientific Computing Center, which is supported by the Office of
Science of the U.S. Department of Energy under Contract No.
DE-AC02-05CH11231.
The work of J. Carlson and S. Gandolfi
were supported by the Department of Energy Nuclear Physics Office,
and by the NUCLEI SciDAC program.
The work of S. Gandolfi was also supported by a Los Alamos LDRD early
career grant.
}


%

\end{document}